\documentclass{moriond}

\usepackage{amsmath,amssymb,amsthm,amsfonts}
\usepackage{siunitx}

\newcommand{\mev}{\text{MeV}}
\newcommand{\gev}{\text{GeV}}
\newcommand{\tev}{\text{TeV}}

\newcommand{\mm}{\text{mm}}

\newcommand{\m}{\text{m}}

\newcommand{\kg}{\text{kg}}

\def\Journal#1#2#3#4{{#1} {\bf #2}, #3 (#4)}

\def\NIMA{{\em Nucl. Instrum. Methods} A}

\def\PLB{{\em Phys. Lett.}  B}
\def\PRL{\em Phys. Rev. Lett.}
\def\PRD{{\em Phys. Rev.} D}

\def\JHEP{{\em JHEP} }
\def\JINST{{\em JINST} }

\def\be{\begin{equation}}
\def\ee{\end{equation}}
\def\bea{\begin{eqnarray}}
\def\eea{\end{eqnarray}}

\newcommand{\Photo}{}

\begin{document}
\vspace*{4cm}
\title{First Physics Results from the FASER Experiment}

\author{ B. Petersen on behalf of the FASER collaboration }

\address{Department of Experimental Physics, CERN,\\
Espl. des Particules 1, 1211 Meyrin, Switzerland}

\maketitle\abstracts{ FASER is a new LHC experiment designed to search
  for light, weakly-interacting particles that are produced in
  proton-proton collisions at the ATLAS interaction point and travel
  in the far-forward direction. The first physics results from the
  initial year of data-taking are presented. A search for dark photons
  decaying to an electron-positron pair found no events, yielding new
  constraints on dark photons with couplings $\epsilon \sim 10^{-5} -
  10^{-4}$ and masses $\sim 10~\mev - 100~\mev$. A search for
  muon-neutrino charged-current interactions in a tungsten target at
  the front of the FASER experiment found $153^{+12}_{-13}$ neutrino
  candidates with a negligible background. The reconstructed
  charge and momentum distributions imply the observation of both
  neutrinos and anti-neutrinos with an incident neutrino energy above 200 GeV.}

\section{Introduction}

FASER~\cite{Feng:2017uoz,FASER:2022hcn} is a new LHC experiment
designed to search for dark photons and other long-lived
beyond-standard model particles as well as to detect high energy
neutrinos produced in proton-proton collisions.  The FASER
detector is located approximately \SI{480}{\meter} from the ATLAS interaction
point (IP1) along the beam collision axis line-of-sight (LOS). Because
the particles are extremely weakly interacting, dark photons and
neutrinos produced at IP1 may travel along the LOS, pass through \SI{100}{\meter}
of rock and concrete without interacting, and then decay or interact
in FASER.  At the same time, most other Standard Model (SM) particles
produced at the ATLAS IP will either be bent away by the LHC magnets
or stopped in the \SI{100}{\meter} of shielding. In these proceedings, the first
results from the search for dark photons~\cite{DarkPhotonConf}
decaying to $e^+ e^-$ and the first direct observation of
neutrinos~\cite{NeutrinoPaper} produced at a particle collider are
presented.

\section{The FASER Detector and Datasets}

A detailed description of the FASER detector can be found
elsewhere.~\cite{FASER:2022hcn} The experiment is located in the TI12
tunnel, which connects the SPS and LHC tunnels, approximately
\SI{480}{\meter} downstream of the ATLAS interaction point. The FASER
detector is partially immersed in a magnetic field and consists of a
passive tungsten/emulsion neutrino detector (FASER$\nu$), two
scintillator-based veto systems, additional scintillators for
triggering, a tracking spectrometer, a pre-shower, and an
electromagnetic calorimeter.

FASER$\nu$ consists of 730 layers of \SI{1.1}{\milli\meter}-thick
\SI{25}{\centi\meter} by \SI{30}{\centi\meter} tungsten plates
interleaved with emulsion films with a total mass of 1.1 metric tons.
The emulsion films are not used in the presented analyses and instead
the FASER$\nu$ detector is used as a target for neutrino interactions
and as an absorber of neutral hadrons produced in muon interactions
upstream of FASER. FASER$\nu$ is situated between two veto
scintillator stations. These are used to veto incoming charged particles, mainly
muons, as well as particles produced in FASER$\nu$ itself in the case
of the dark photon search. The front veto station has two layers of
$\SI{30}{\centi\meter} \times \SI{35}{\centi\meter}$,
\SI{2}{\centi\meter}-thick plastic scintillators, while the second
veto station has three layers of $\SI{30}{\centi\meter} \times
\SI{30}{\centi\meter}$, \SI{2}{\centi\meter}-thick plastic
scintillators. All scintillators are read out with photomultiplier
tubes (PMTs).

The tracking system consists of the interface tracking station (IFT)
and the three tracking spectrometer
stations.~\cite{FASER:2021ljd} Each tracking station is composed of
three planes with eight ATLAS semiconductor tracker (SCT) barrel
modules~\cite{Abdesselam:2006wt} per plane, arranged as two columns of
four modules covering \SI{24}{\centi\meter} by \SI{24}{\centi\meter}.
%Each SCT module consists of a double-layer of
%single-sided silicon microstrips with a \SI{40}{\milli\radian} stereo
%angle and an \SI{80}{\um} strip pitch.
In the presented analyses, only
the spectrometer is used to reconstruct charged particles, while the
IFT is only used to measure remnants of the neutrino interactions
through cluster counting.

Between the three tracking spectrometer stations are two
\SI{1}{\meter}-long dipole magnets with magnetic field of 0.57~T.  A
similar \SI{1.5}{\meter}-long magnet is placed between the last veto
station and the spectrometer and serves as the decay volume for dark
photon decays. All magnets have a circular aperture of 200 mm diameter, which
defines the active transverse area of the detector, and bend charged
particles in the vertical plane. In addition, the so-called timing
scintillator station is located between the decay volume and the front
of the first tracking station of the spectrometer. It is used
primarily for triggering.

The electromagnetic energy of particles in the detector volume is
measured by the electromagnetic calorimeter, located as the most
downstream component of the detector. The calorimeter is made up of
four spare modules from the LHCb experiments outer ECAL
calorimeter.~\cite{LHCB:2000ab} Each module is $\SI{12}{\centi\meter}
\times \SI{12}{\centi\meter}$ in the transverse plane and the modules
are arranged in a two-by-two formation.
%, and is made up of 66 layers of interleaved
%2~mm thick plastic scintillator and 1~mm thick lead plates,
%corresponding to a total of 25 radiation lengths.
The calorimeter modules are readout using PMTs. In order to maintain good PMT
linearity and avoid saturation for deposits above a few hundred GeV,
optical filters were installed in front of the PMTs for the second
part of the data-taking period in 2022. Between the calorimeter and
the tracking spectrometer is a pre-shower system built from two
scintillator layers of the same type as used for the second veto
station, each preceded by a \SI{3}{\mm}-thick tungsten absorber plate.

The trigger and data acquisition system of FASER was designed to
achieve high efficiency and reliability.~\cite{FASER:2021cpr} Neutrino
and dark photon candidate events are triggered by scintillator signals
that exceed a preset threshold below that of a single minimum-ionizing
particle (MIP), resulting in a typical trigger rate of
0.5-\SI{1.3}{\kilo\Hz}. Data was recorded throughout 2022 with
an integrated data taking efficiency of 96.1\%.

The presented analyses use data from runs with stable beam
conditions collected between July (September) and November 2022 for
the neutrino (dark photon) analysis. The latter analysis requires the
calorimeter to have the optical filters installed. This corresponds to
total luminosities as measured by ATLAS of \SI{35.4}{\per\femto\barn} and
\SI{27.0}{\per\femto\barn}, respectively after data quality
selection. Event reconstruction is carried out in FASER's
Calypso offline software system, based on the
open-source Athena framework from
the ATLAS experiment.

Monte Carlo (MC) simulation samples are used to evaluate the signal
efficiency and in the estimation of some of the backgrounds, and
systematic uncertainties. All samples are simulated using
\texttt{GEANT4}~\cite{G4} with a detailed description of the detector
geometry, including passive material. The samples include a realistic
level of detector noise, and are reconstructed in the same way as the
data.

\section{Search for Dark Photon Decays}

Dark photons are a common feature of hidden sector models motivated by
dark matter.  If the dark sector contains a U(1) electromagnetic
force, the dark gauge boson can mix with the SM gauge
boson,~\cite{Holdom:1985ag} leading to a new spin-1 gauge boson, the
dark photon $A'$, which is among the leading targets for dark sector
searches. The dark photon is characterized by its mass, $m_{A'}$, and
coupling parameter, $\epsilon$. FASER is most sensitive to dark
photons with $m_{A'} \sim 10-100~\mev$ and $\epsilon \sim 10^{-5}$--$10^{-4}$.
In this region of parameter space, the dark
photon is primarily produced at the LHC through neutral pion decays to
$A'\gamma$, but $\eta\rightarrow A'\gamma$ decays and dark
bremsstrahlung $pp\rightarrow ppA'$ also contribute. In the forward
direction, the produced dark photons are highly energetic which,
depending on the coupling and mass, can result in a long decay
length. For $E_{A'} \gg m_{A'} \gg m_e$, the dark photon decay length
can be expressed as~\cite{Feng:2017uoz}
\begin{equation}
\label{eq:ap_decay_length}
L = c \beta \tau \gamma  \approx (80~\m ) \left[ \frac{10^{-5}}{\epsilon} \right]^2 
\left[ \frac{E_{A'}}{\tev} \right] \left[ \frac{100~\mev}{m_{A'}} \right]^2 \ .
\end{equation}
This means a substantial number of these dark photons can decay at the
FASER location. For $\SI{1}{\mev} <m_{A'}<\SI{211}{\mev}$, the dark
photon will decay 100\% to an $e^{+}e^{-}$ pair. The analysis
therefore searches for the appearance of a track pair in the decay
volume before the spectrometer with a large energy deposition in the
electromagnetic calorimeter.

\begin{figure}[tbp]
\includegraphics[width=0.49\textwidth]{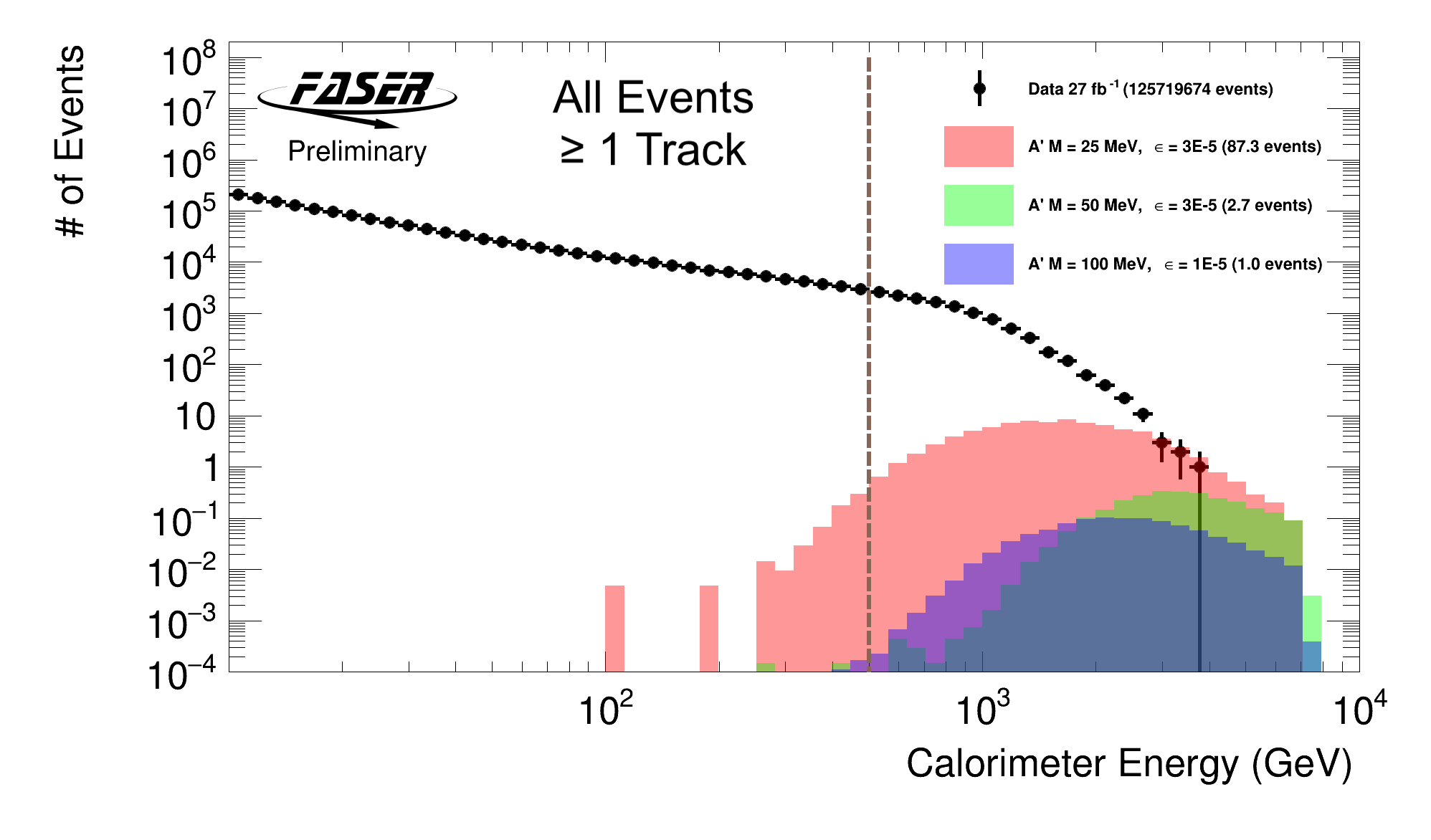}
\includegraphics[width=0.49\textwidth]{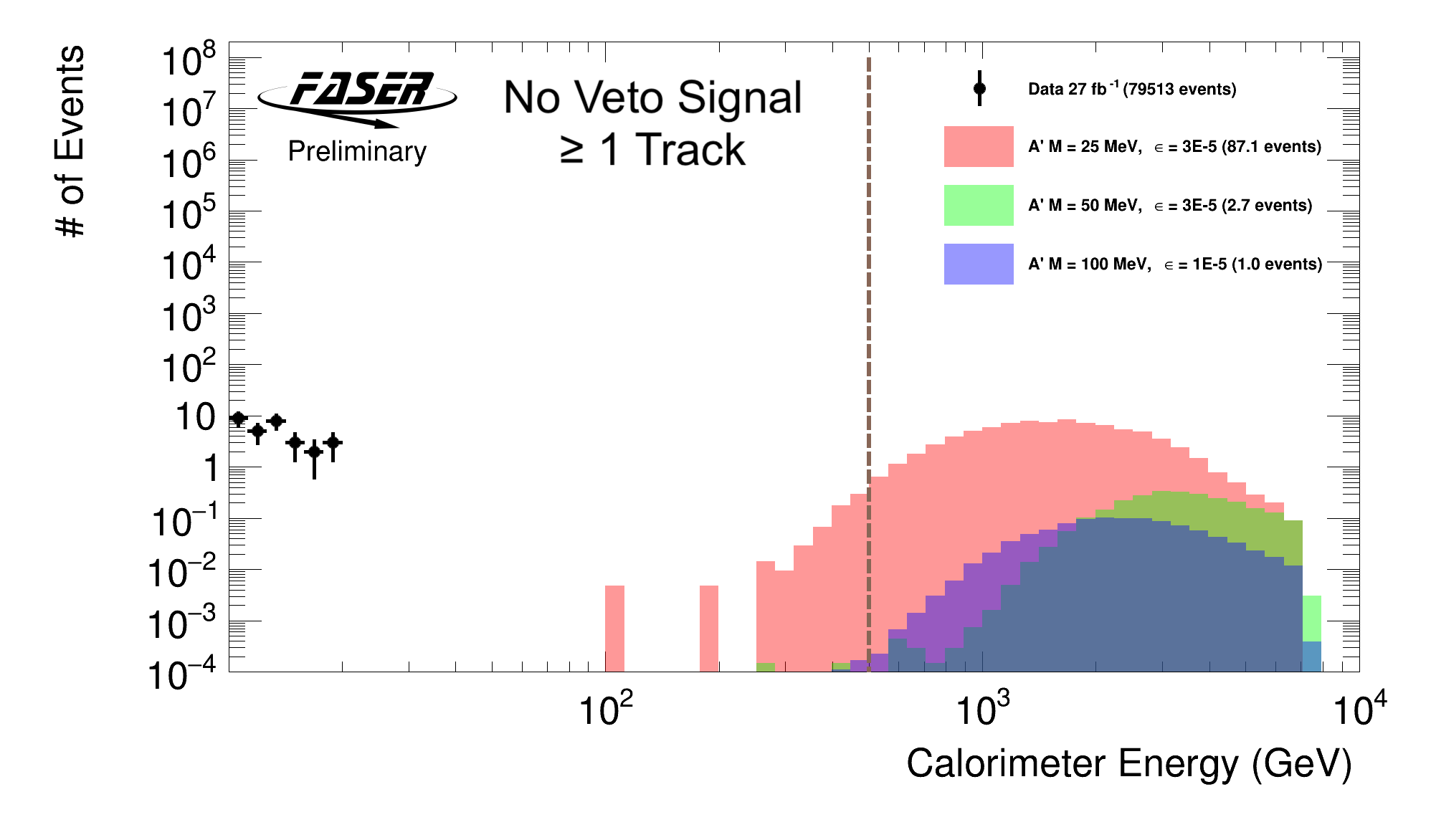}
\includegraphics[width=0.49\textwidth]{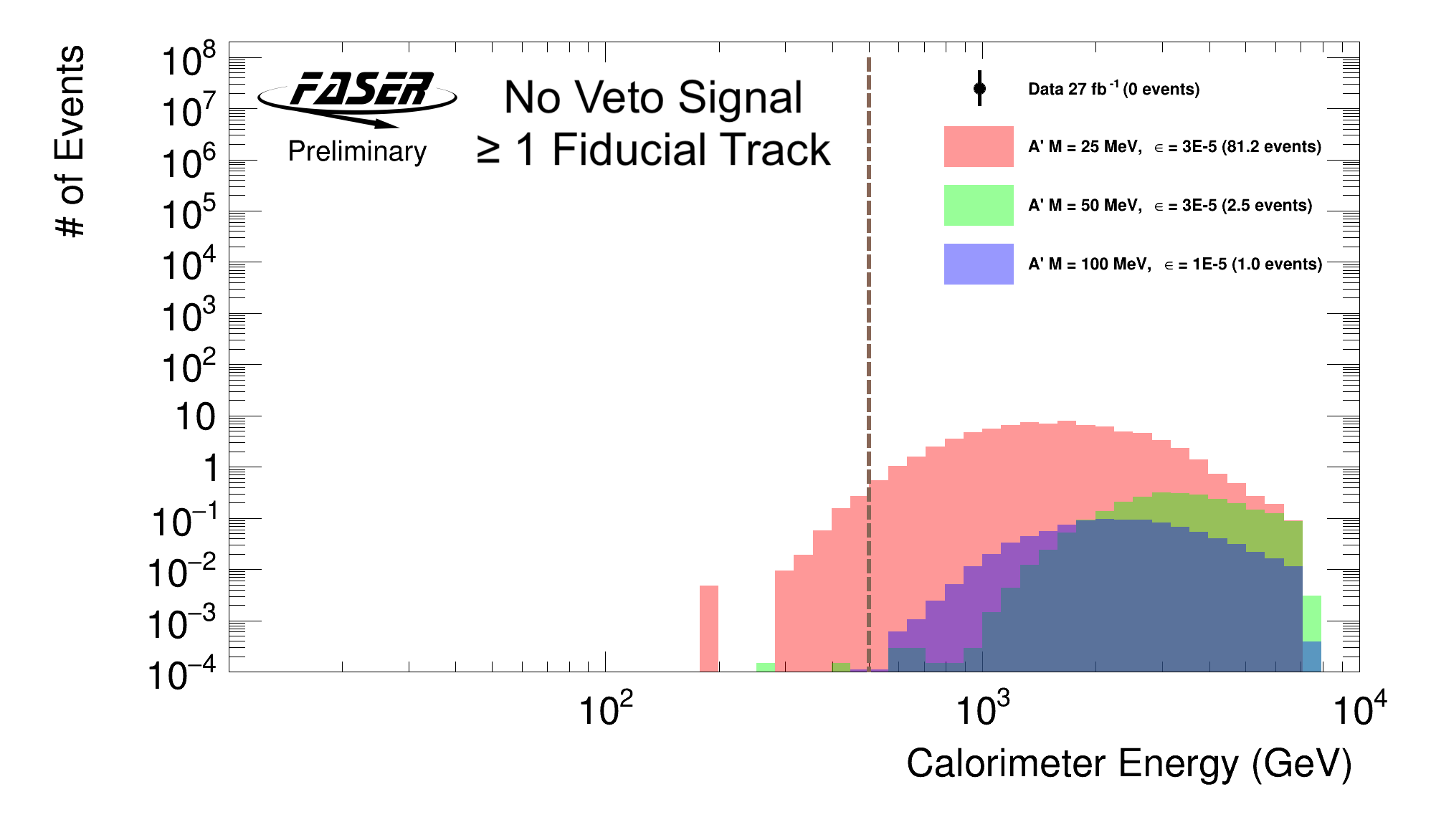}
\includegraphics[width=0.49\textwidth]{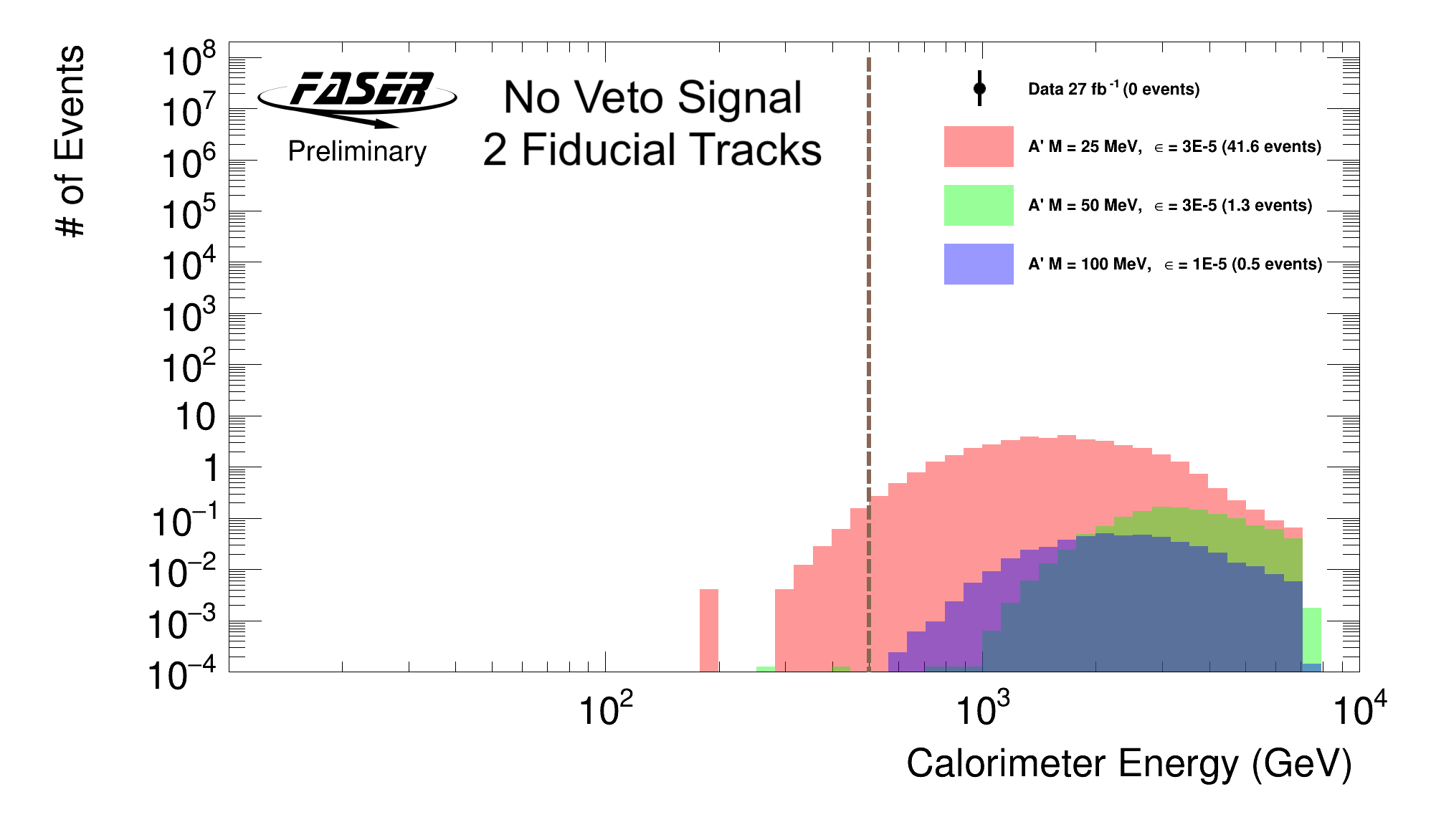}
\caption{The calorimeter energy distribution after various levels of selections for data and expected signal for three representative signal models. (top left) after a reconstructed track requirement; (top right) after a reconstructed track requirement, and with no signal in the veto scintillators; (bottom left) after a fiducial track requirement and no veto scintillator signal; (bottom right) after a fiducial two track requirement and no veto scintillator signal.}   
\label{fig:selectionPlots}
\end{figure}

The search uses simple, robust selection criteria optimized for an
early discovery with selections fixed before looking at data in the
signal region.  All five veto scintillator layers are required to have
no signal, vetoing events with any incoming charged particles or
interactions in the FASER$\nu$ detector. The timing and pre-shower
scintillators layer are required to have signals consistent with two
or more MIPs. Exactly two good quality reconstructed tracks with
momentum above 20 GeV inside a fiducial radius of $\SI{95}{\mm}$ are
required.  The extrapolations of both tracks to the first scintillator
station are required to have a radius within $\SI{95}{\mm}$ in order
to suppress particles coming at a large angle. Finally at least
\SI{500}{\gev} of deposited energy in the calorimeter is required.
This selection results in a typical signal efficiency of 40-50\% for
dark photons decaying inside the decay volume.

The main backgrounds remaining after all selections are estimated to
be from neutrino interactions in the detector and neutral hadrons
entering the detector and decaying in the decay volume. The former is
estimated from large simulation samples to be $0.0018\pm0.0024$ events
where most of these originate from neutrino interactions in the timing
scintillator. Neutral hadrons originate from muon interactions in the
rock upstream of FASER and are mostly suppressed by the detection of
the muon and absorption of the hadrons in the FASER$\nu$ tungsten. The
background is estimated to be $(2.2\pm3.1)\times10^{-4}$ events based
on low energy events with two and three reconstructed tracks with
different veto scintillator conditions.  Background events from
undetected muons are negligible as each of the five scintillator
layers have a MIP-detection efficiency measured in data to be more than
$99.9998\%$. Similarly, backgrounds from cosmic ray events and
non-collision beam backgrounds have been studied
%using non-colliding bunches and data recorded without beams
and found to be negligible.

\begin{figure}[tbp]
  \centering
\includegraphics[width=0.7\textwidth,trim=0 0 0 80,clip]{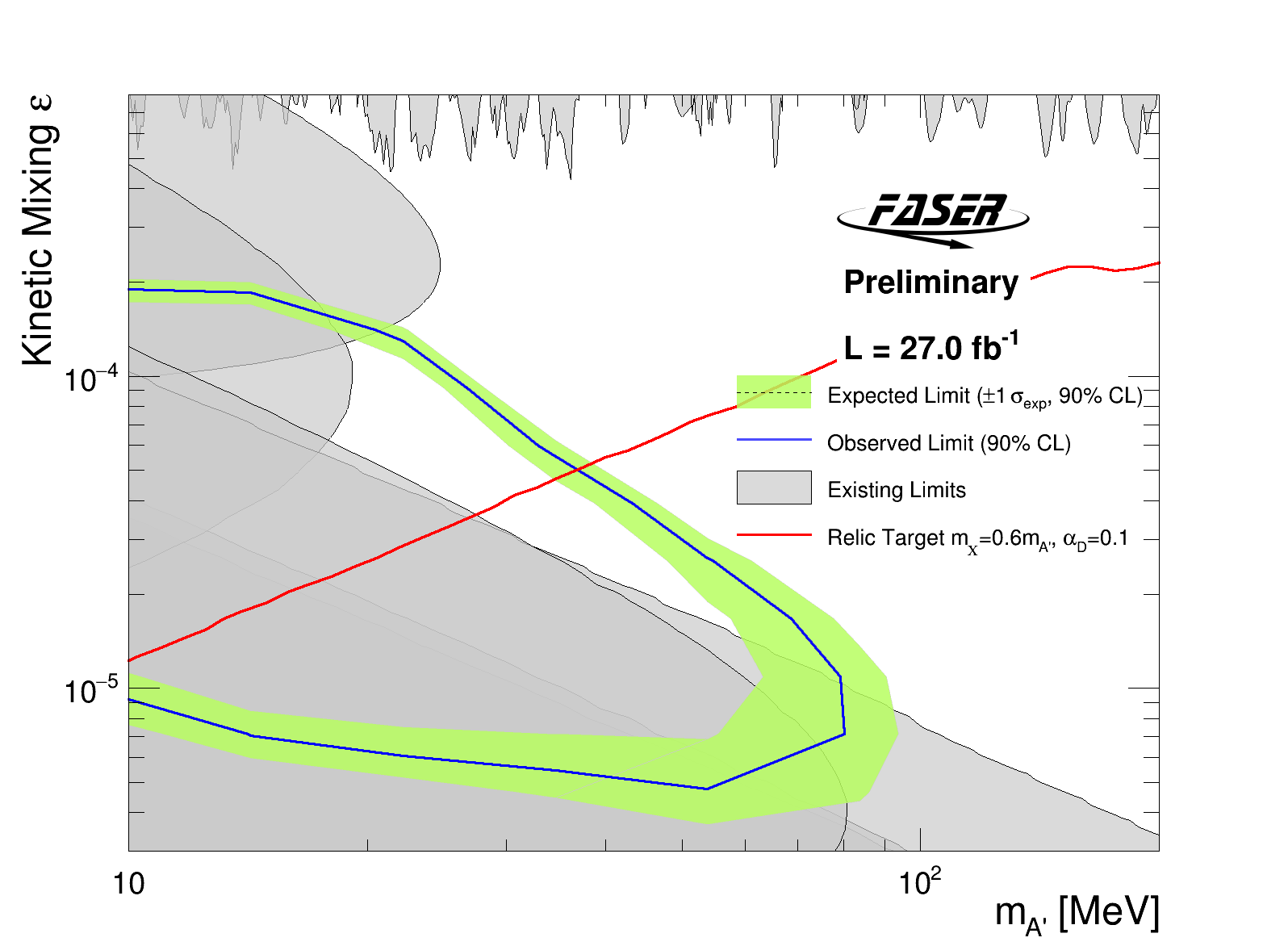}
\caption{The 90\% confidence level exclusion contour in the dark photon model parameter space. Regions excluded by previous experiments are shown in grey\protect\cite{Ilten:2018crw}.
  The red line shows the region of parameter space which yields the correct dark matter relic density, with the assumptions discussed in the main text.} 
\label{fig:exclusion-limit}
\end{figure}

Figure~\ref{fig:selectionPlots} shows the calorimeter energy
distribution after different stages in the event selection. No data
events remain after just requiring no signal in the veto layers and at
least one track in the fiducial region. Given that no events are
observed, the results are used to set exclusion limits on dark photons
at 90\% confidence level.  Figure~\ref{fig:exclusion-limit} shows the
exclusion limit in the signal parameter space. The analysis excludes
signal models in the range $\epsilon \sim 1 \times 10^{-5} - 2 \times
10^{-4}$ and masses $\sim 10~\mev - 80~\mev$, and provides exclusion
for previously viable models in the range $\epsilon \sim 2 \times
10^{-5} - 1 \times 10^{-4}$ and masses $\sim 17~\mev - 70~\mev$.
Figure~\ref{fig:exclusion-limit} also includes an example thermal
relic contour~\cite{Kling:2021fwx}, obtained for the scenario where
the dark photons couple to a light complex scalar dark matter field
$\chi$, demonstrating that FASER is sensitive to dark photons in a
cosmologically interesting region of parameter space.

\section{Observation of Neutrino Interactions}

Until now no neutrino produced at a particle collider has ever been
directly detected. Colliders copiously produce both neutrinos and
anti-neutrinos of all flavors, and they do so in a range of very high
energies where neutrino interactions have not yet been observed.
Nevertheless, collider neutrinos have escaped detection, because they
interact extremely weakly, and the highest energy neutrinos, which
have the largest probability of interacting, are predominantly produced
in the forward region, parallel to the beam line. In 2021, the FASER
collaboration identified the first collider neutrino
candidates~\cite{FASER:2021mtu} using a \SI{29}{\kg} pilot detector,
highlighting the potential of discovering collider neutrinos in LHC
collisions.

The presented search targets $\nu_\mu$ and $\overline \nu_\mu$
charged-current interactions in the tungsten/emulsion detector. Such
interactions will produce a high-momentum muon that can be
reconstructed in the three stations of the FASER tracking spectrometer
as well as increased activity in the second veto scintillator station
and in the IFT tracking station from secondary particles produced in
the interaction. The event selection starts from events triggered by
any of the scintillators downstream of FASER$\nu$ that are consistent
with a colliding bunch crossing identifier and all downstream
scintillator layers are required to have a signal consistent with at
least one MIP.  Events with signals in any of the two layers of the
first veto station are rejected as they are likely from incoming muons
or other charged particle backgrounds.  Only events with a single,
good quality track in the three tracker stations with a reconstructed
track momentum above \SI{100}{GeV} are accepted.  The track is
required to be in the fiducial tracking volume
($r<\SI{95}{\milli\meter}$) of the spectrometer and when extrapolated
to the IFT. It should be within $r<\SI{120}{\milli\meter}$ at the
first veto scintillator.
%and to have a polar angle below \SI{25}{\milli\radian}.
Based on simulation of neutrino production from proton-proton
collisions~\cite{Kling:2021gos} and of interactions~\cite{Genie2010} in FASER$\nu$, $151\pm41$ neutrino
events are expected to be selected with the above selection. The
uncertainty is given by the difference between two different
production models and no experimental uncertainties are included in this
estimate.

\begin{figure}[tbp]
  \centering
  \includegraphics[width=0.47\linewidth]{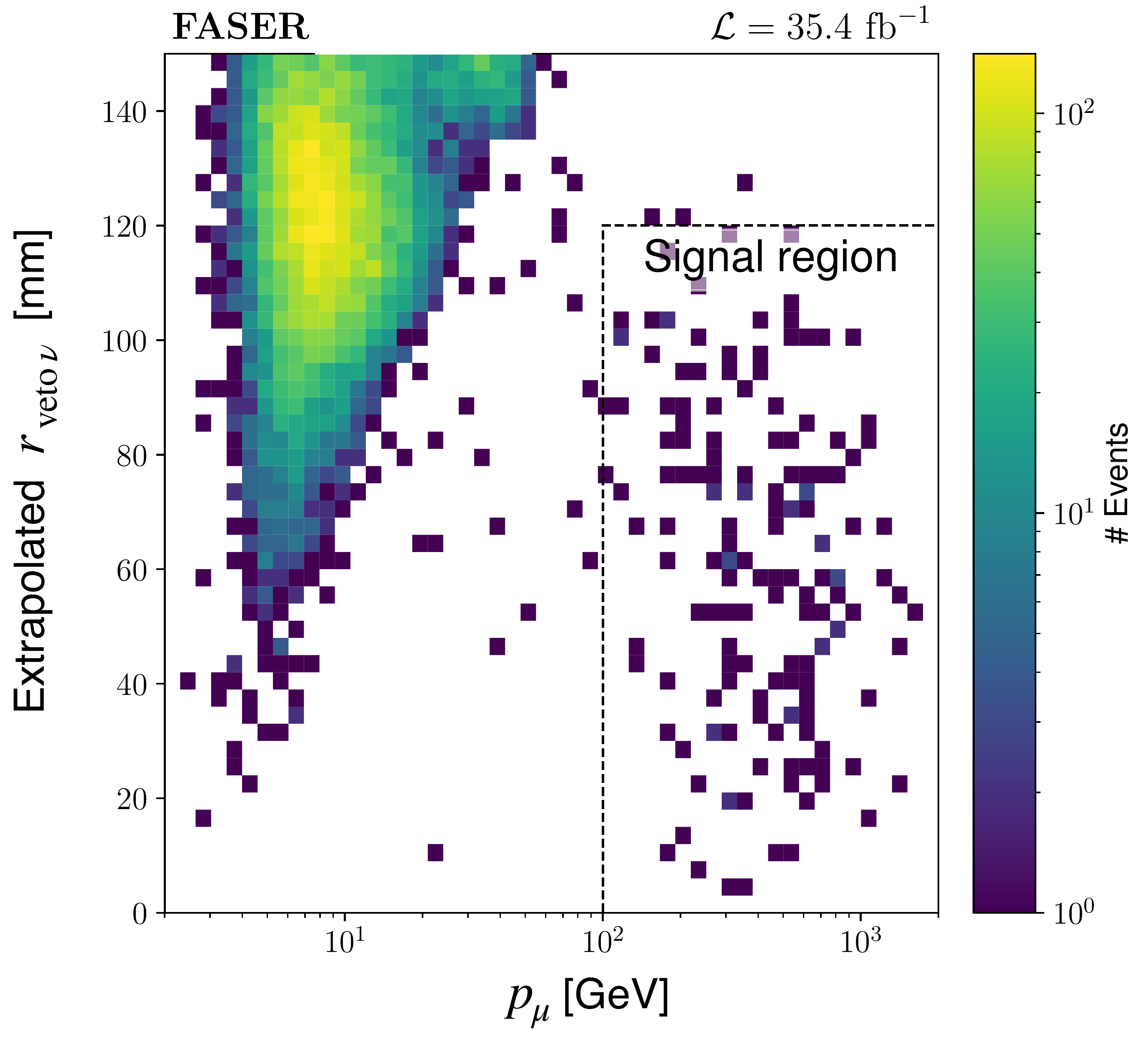}
  \includegraphics[width=0.44\linewidth]{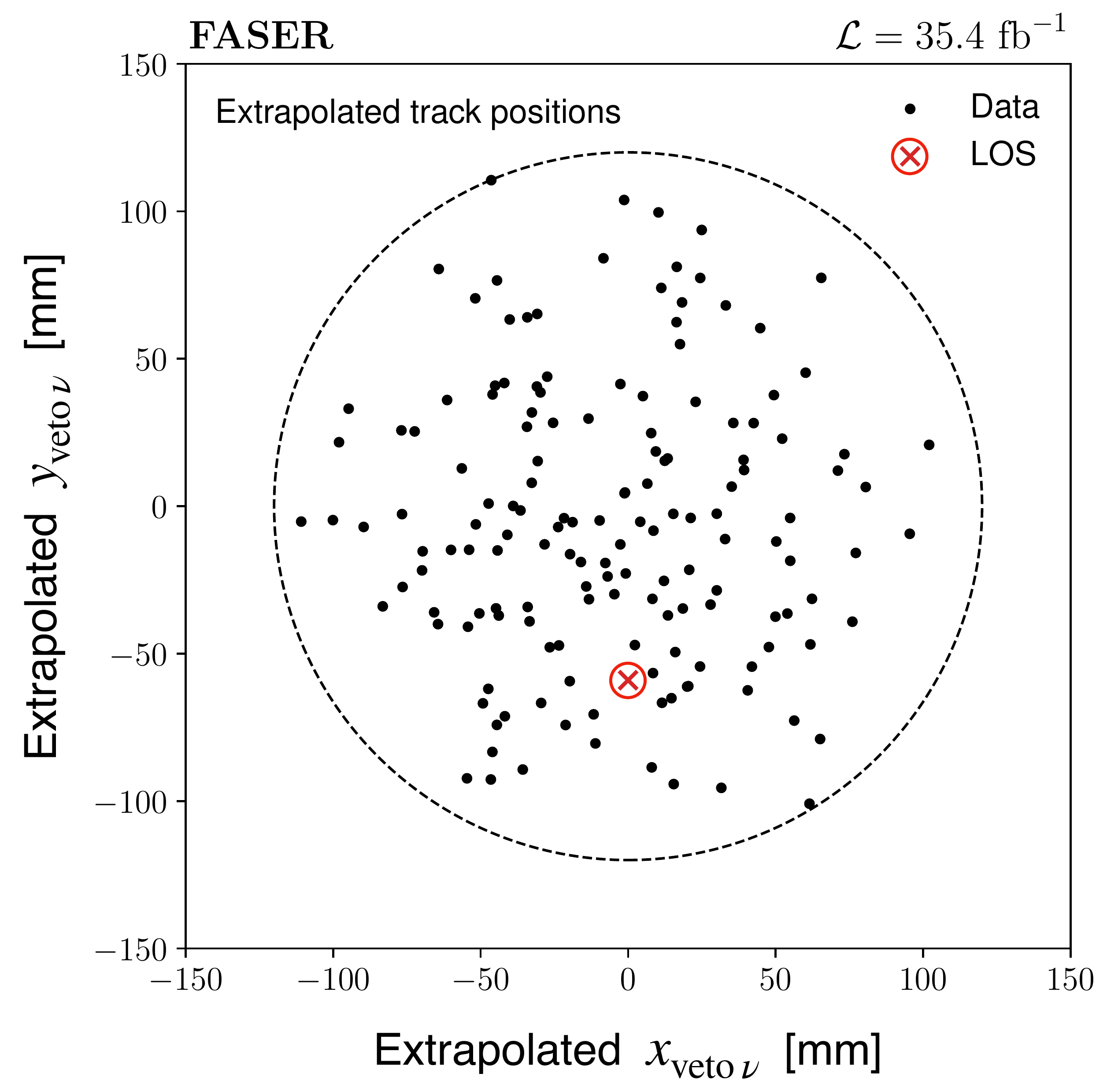}
  \caption{ Distribution of neutrino candidate tracks extrapolated to the first veto station. (left) The
    extrapolated track radius $r_{\mathrm{veto}\, \nu}$ and
    reconstructed track momentum $p_\mu$ is depicted including outside
    the final signal region.  The region with lower momenta and larger
    radii is dominated by background events consisting of charged
    particles that miss the FASER$\nu$ scintillator station.  (right)
    extrapolated transverse position of the reconstructed tracks of
    neutrino-like events to the FASER$\nu$ scintillator station. The
    ATLAS LOS is indicated with a red marker.  }
\label{fig:sel_events}  
\end{figure}

The background from neutral hadrons is estimated from simulation of
neutral hadron production from muon interactions in the rock in front
of the FASER detector combined with simulation of neutral hadron
interactions in FASER$\nu$ that produce charged particles with a
momentum of more than \SI{100}{\gev}. A total of just $0.11\pm0.06$
events are expected. An additional background can come from large
angle muons that miss the front veto station and scatter into the
spectrometer in FASER$\nu$. This is estimated to be $0.08 \pm 1.83$
events based on the rate of events in data within a large-radius
control region.

\begin{figure}[tbp]
  \includegraphics[width=0.33\linewidth]{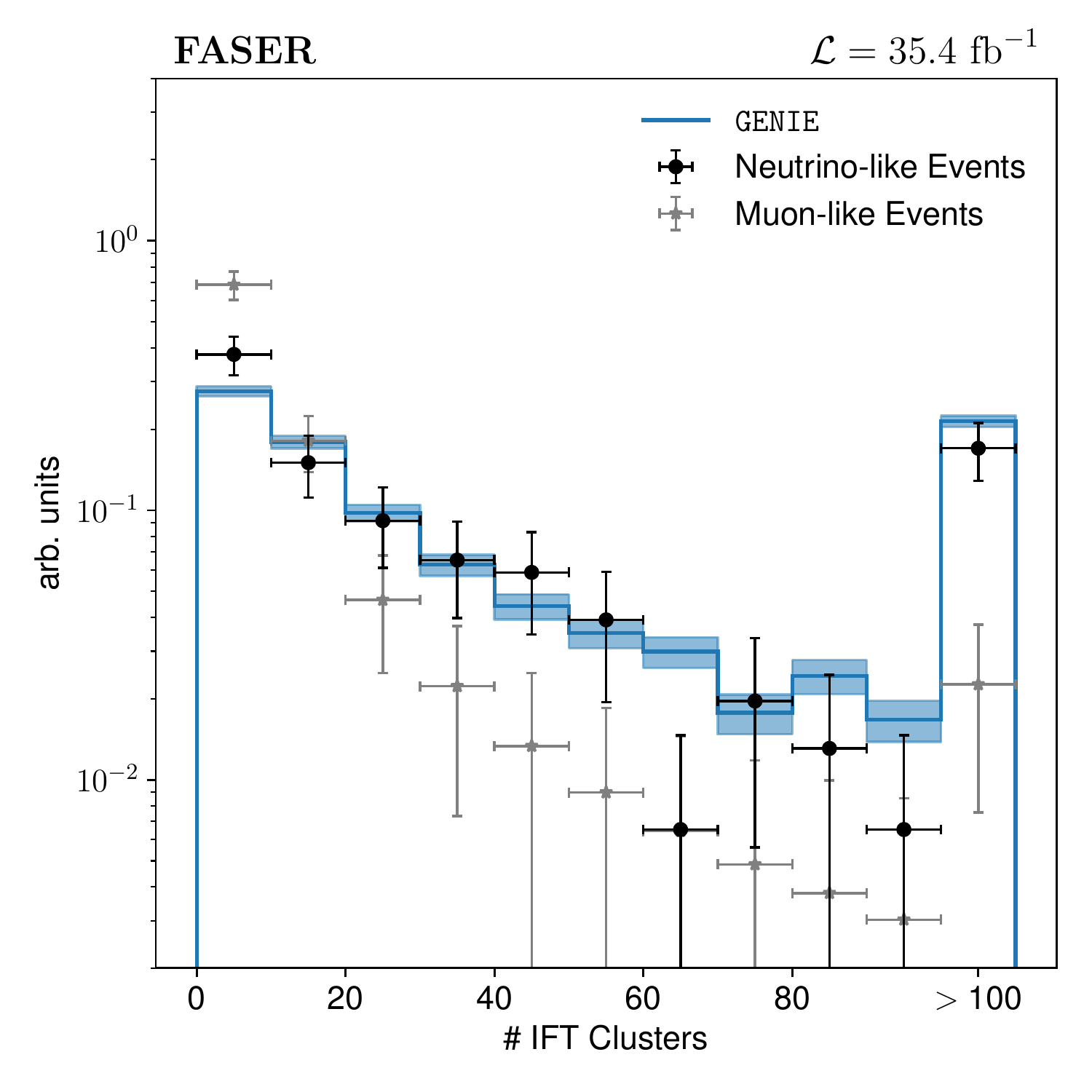}
  \includegraphics[width=0.33\linewidth]{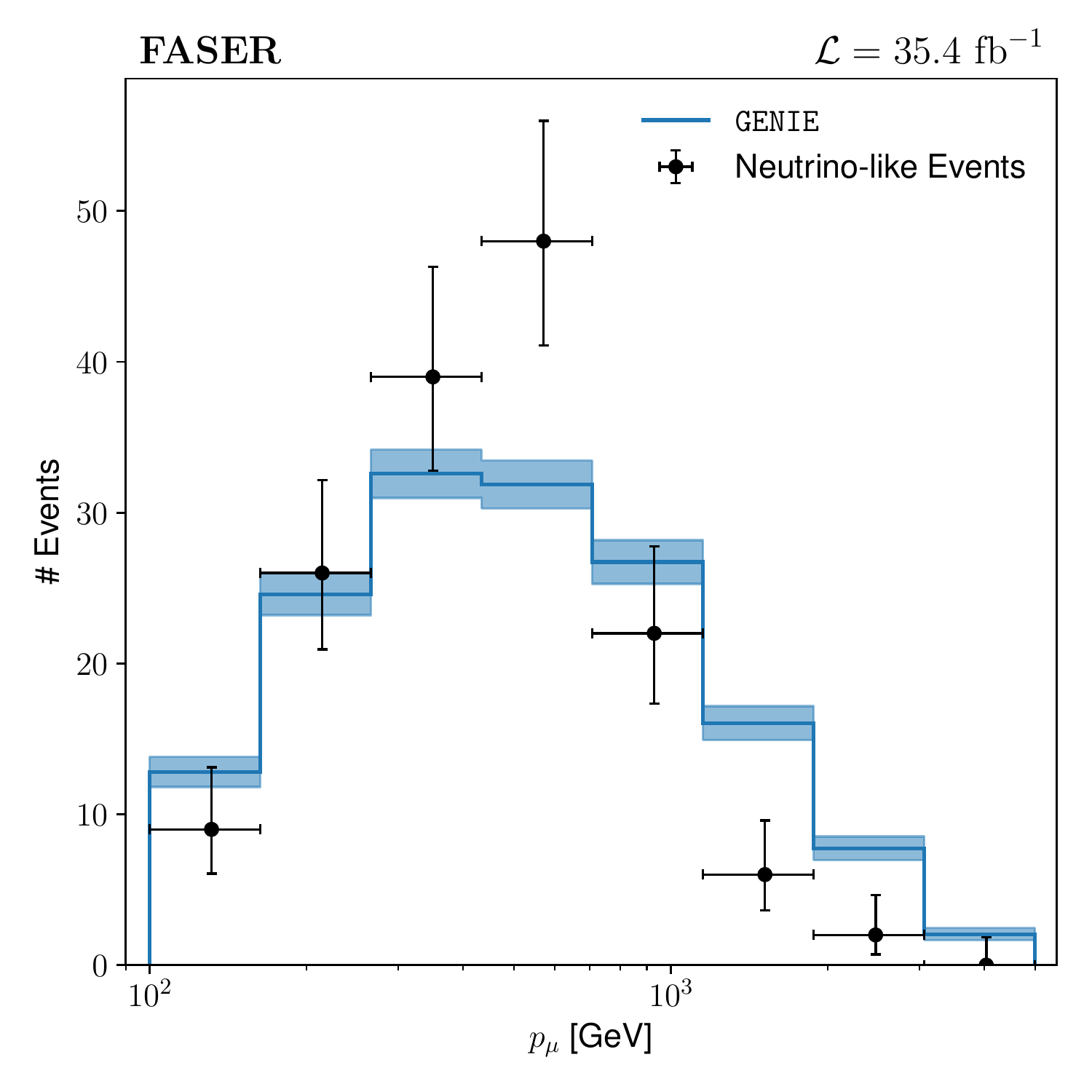}  
  \includegraphics[width=0.33\linewidth]{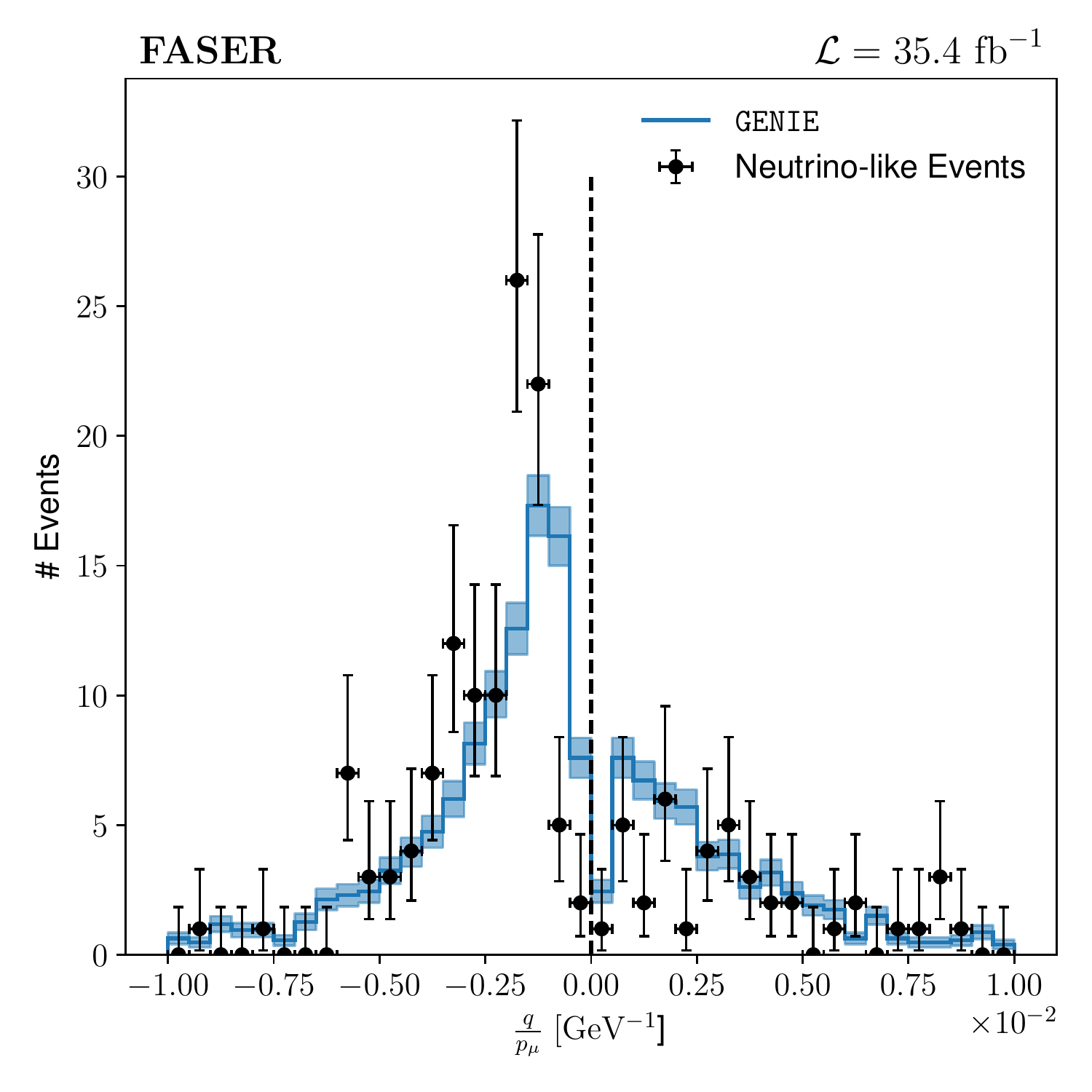}    
  \caption{The figures depict the number of reconstructed clusters in
    the IFT, momentum, $p_\mu$ and $q/p_\mu$ for events in the signal
    region (black markers) and compares them to the expectation from
    simulation \texttt{GENIE} (blue) and muon-like events (grey
    markers). The muon-like events are from events where both layers
    of the front scintillator station observed a signal, and show the
    expected distributions for non-neutrino backgrounds. The blue
    bands correspond to the statistical error of the simulated samples
    and are luminosity scaled for $p_\mu$ and $q/p_\mu$ and normalized to unity
    for clusters.  }
\label{fig:IFT_cluster_q_over_p_momentum}       
\end{figure}

Figure~\ref{fig:sel_events} shows the selected events, as well as
background-enriched regions around the signal region. In total we
observe 153 events passing all selection steps. The number of neutrino
events and the statistical significance are estimated using a binned
extended maximum likelihood fit taking into account the background
estimates above and fitting as well the number of events with signals
in one or both of the veto layers to account for the number of incoming
muons that fail to be detected by the front veto scintillator
station. The fitted number of neutrino events is $153^{+12}_{-13}$
events with a statistical significance of 16 standard deviations over
the background-only hypothesis. The right side of
Figure~\ref{fig:sel_events} shows that the events are distributed
around the LOS as expected. Additional properties of the neutrino
events are displayed in
Figure~\ref{fig:IFT_cluster_q_over_p_momentum}. It confirms that the
events have larger activity in the IFT than incoming muon events and
that the reconstructed muons and consequently also the interacting
neutrinos have energies that are significantly larger than
\SI{200}{\GeV}.  A clear charge separation in $q/p_\mu$ for the
reconstructed tracks is also observed with a total of 40 events with a
positively-charged track, demonstrating the presence of anti-neutrinos
in the analyzed data set.

\section{Summary}

The first physics results from the FASER experiment have been
presented.  No signal events are observed in an almost background free
search for dark photon decays to $e^{+}e^{-}$ and new limits are set
in a cosmological relevant region of parameter space. FASER has
performed the first direct detection of neutrinos produced at a
collider~\footnote{In addition to FASER, the SND@LHC experiment
  reported evidence of neutrino interactions elsewhere in these
  proceedings.} with $153^{+12}_{-13}$ neutrino events observed from
charged-current interactions of $\nu_\mu$ and $\overline \nu_\mu$ in
FASER$\nu$. These results represent the beginning of the FASER physics
programme with many more results to come in the next years with up to
an order of magnitude more luminosity expected.

%\begin{table}[t]
%\caption[]{Experimental Data bearing on $\Gamma(K \ra \pi \pi \gamma)$
%for the $\ko_S, \ko_L$ and $K^-$ mesons.}
%\label{tab:exp}
%\vspace{0.4cm}
%\begin{center}
%\begin{tabular}{|c|c|c|l|}
%\hline
%& & & \\
%&
%$\Gamma(\pi^- \pi^0)\; s^{-1}$ &
%$\Gamma(\pi^- \pi^0 \gamma)\; s^{-1}$ &
%\\ \hline
%\mco{2}{|c|}{Process for Decay} & & \\
%\cline{1-2}
%$K^-$ &
%$1.711 \times 10^7$ &
%\begin{minipage}{1in}
%$2.22 \times 10^4$ \\ (DE $ 1.46 \times 10^3)$
%\end{minipage} &
%\begin{minipage}{1.5in}
%No (IB)-E1 interference seen but data shows excess events relative to IB over
%$E^{\ast}_{\gamma} = 80$ to $100MeV$
%\end{minipage} \\
%& & &  \\ \hline
%\end{tabular}
%\end{center}
%\end{table}

%\section*{Acknowledgments}

%This is where one places acknowledgments for funding bodies etc.

\section*{References}

%\bibliographystyle{unsrt}
%\bibliographystyle{utphys}    
%\bibliography{faser}

\end{document}